\documentclass[%
 reprint,
 showkeys,
 amsmath,amssymb,
 aps,
 pre,
]{revtex4-1}

\usepackage{graphicx}
\usepackage{dcolumn}
\usepackage{bm}
\usepackage[percent]{overpic}
\usepackage{hyperref}

\newcommand{\ignore}[1]{}

\graphicspath{ {figs/} }

\begin{document}

\preprint{APS/123-QED}

\title{Evacuation time estimate for a total pedestrian evacuation using \\ queuing network model and volunteered geographic information}

\author{Bharat Kunwar}
 \email{b.kunwar@bristol.ac.uk}

\author{Filippo Simini}%
 \email{filippo.simini@bristol.ac.uk}

\author{Anders Johansson}%
 \email{a.johansson@bristol.ac.uk}

\affiliation{Faculty of Engineering, University of Bristol.}

\date{\today}

\begin{abstract}





Estimating city evacuation time is a non-trivial problem due to the interaction between thousands of individual agents, giving rise to various collective phenomena, such as bottleneck formation, intermittent flow and stop-and-go waves. We present a mean field approach to draw relationships between road network spatial attributes, number of evacuees and resultant evacuation time estimate (ETE). We divide $50$ medium sized UK cities into a total of $697$ catchment areas which we define as an area where all agents share the same nearest exit node. In these catchment areas, 90\% of agents are within $5.4$ km of their designated exit node. We establish a characteristic flow rate from catchment area attributes (population, distance to exit node and exit node width) and a mean flow rate in free-flow regime by simulating total evacuations using an agent based `queuing network' model. We use these variables to determine a relationship between catchment area attributes and resultant ETE. This relationship could enable emergency planners to make rapid appraisal of evacuation strategies and help support decisions in the run up to a crisis.

\end{abstract}

\pacs{Valid PACS appear here}
\maketitle


\section{\label{sec:introduction}Introduction}

Interaction between individual agents, city topology, disaster type, evacuation mode, information propagation patterns and stochastic variables can all influence the temporal extent of a city-wide evacuation. Additionally, growing urban populations \citep{desa2013world} amplifies the impact of extreme events \citep{MunichRE}. There is a need to examine factors affecting evacuation time in relation to the latest understanding of crowd dynamics and evacuation behaviour.


Evacuation time estimate (ETE) analysis (a) tells emergency planners if an evacuation plan can reduce hazard exposure time (b) measures effect of uncontrollable events such as adverse weather and (c) assesses whether traffic management actions help reduce it \cite{UrbanikII2000}.
A study of flood evacuation in Netherlands identifies a need for alternative evacuation strategies for coastal areas after it found that it was not feasible to evacuate preventively within a 48 hour warning window \cite{kolen2012time}.
EMBLEM2, an empirical study, categorises research findings about evacuees' behaviour in hurricanes into 4 evacuation route system parameters, 16 behavioural parameters and 5 evacuation scope/timing parameters to calculate ETE \cite{lindell2008emblem2}.
A sensitivity analysis of radiological emergency micro traffic simulation finds that ETE is sensitive to traffic factors (interaction with pedestrians, intersection traversing time, car ownership, etc.) and route choice mechanisms (shortest path and myopic behaviour) \cite{sinuany1993simulating}.
NETVACl, a macro traffic simulation finds that ETE for areas surrounding nuclear power plant sites are sensitive to road network topology, intersection design and control, and a wide array of evacuation management strategies \cite{Sheffi1982}.
Another study produces ETE for 10 miles radius of 52 nuclear power stations taking consideration of factors such as population density, weather conditions, warning time, response time and confirmation time \cite{urbanik1981analysis}. 

Some models take a dynamic network flow approach to minimise evacuation time \cite{Bretschneider2011,chiu2007nonotice} while others use social force based models like EPES to establish optimal earthquake evacuation behaviour \cite{DOrazio2014}.
The `Last-Mile' project uses a `queuing network' model to obtain an optimal evacuation plan for the Indonesian city of Padang using time-dependent network attributes to imitate conditions of a tsunami \cite{Lammel2010representation}. The underlying flow model simulates traffic taking only free speed, bottleneck capacities and space constraints into account. 
This approach is preceded by an early evacuation plan optimisation study for Yokosuda city in Japan which uses a combination of the shortest path algorithm and minimal cost flow approach accounting for the capacity limit of each place of refuge \cite{yamada1996network}.

Evacuees' behaviour plays an import role during evacuations.
Combination of individual traits and basic social psychological processes such as (a) risk perception, (b) social influence and (c) access to resources predict evacuation behaviour while some population subgroups choose not to evacuate depending on the severity of storm, territoriality, etc. according to a study conducted after Hurricanes Hugo and Andrew \cite{riad1999predicting}. 
Subjective perception of how bad the storm is going to be and the severity of damage also seem to play an important role in evacuation likelihood following a warning \cite{baker1979predicting}. 
The effect of compliance behaviour on ETE has been studied using the EVAQ evacuation model and a case study of the Rotterdam metropolitan area in Netherlands \cite{Pel2010b}.


Crowd dynamics is an important feature in large cities and understanding it is a crucial component of emergency evacuation modelling where Agent Based Modelling (ABM) is increasingly being used for large scale simulations to account for many interacting entities \cite{Johansson2012}.
The transition between low and high density phases are common in social systems like cities \cite{bukavcek2014experimental}.
Keeping a constant lower limit on the net-time headway has been shown as one of the key mechanisms behind emergent crowd dynamics \cite{johansson2009constant}.
Observed collective phenomena in pedestrian crowds include lane formation in corridors and oscillations at bottlenecks in normal situations as well as different kinds of blocked states produced in panic situations \cite{helbing2002simulation}. 
Video recordings of the crowd disaster in Mina/Makkah during the Hajj on January 12, 2006 reveal two subsequent, sudden transitions from laminar to stop-and-go \cite{helbing2006intermittent} and “turbulent” flows \cite{helbing2007dynamics}.
The  transition to turbulent flow is responsible for sudden eruptions of pressure release comparable to earthquakes, which cause sudden displacements and the falling and trampling of people \cite{helbing2000simulating}.
However, from a macroscopic viewpoint, pedestrian behaviour can be assimilated into a relationship between walking speed $v$ and local density $k$, variables familiar to the transport research community \cite{weidmann1993transporttechnik}.

Review of existing work highlights a gap in understanding which relates ETE to interaction between city population and their topological make-up. Topologies can vary between parts of cities, one city to another and one region of the planet to another, all growing in complexity at the same time. A `queuing network' ABM which incorporates pedestrian behaviour and network topology has the potential to define a direct relationship between city topological attributes and their ETE.

\section{\label{sec:method} Methodology}

We will now describe a model used for deriving the necessary quantities required for our analysis. We make the following assumptions across the model:

\begin{itemize}
    \item Evacuation type is a total evacuation scenario, for which exit nodes lie at intersection between major roads and city administrative boundary \cite{kunwar2014large}.
    \item Evacuation mode is by walking only.
    \item Route to exit node is calculated using Dijkstra's shortest path algorithm \cite{dijkstra1959note} (no dynamic routing due to bottlenecks congestion).
\end{itemize}

\begin{figure}
    [!t] \centering

    \begin{overpic}[scale=0.40]{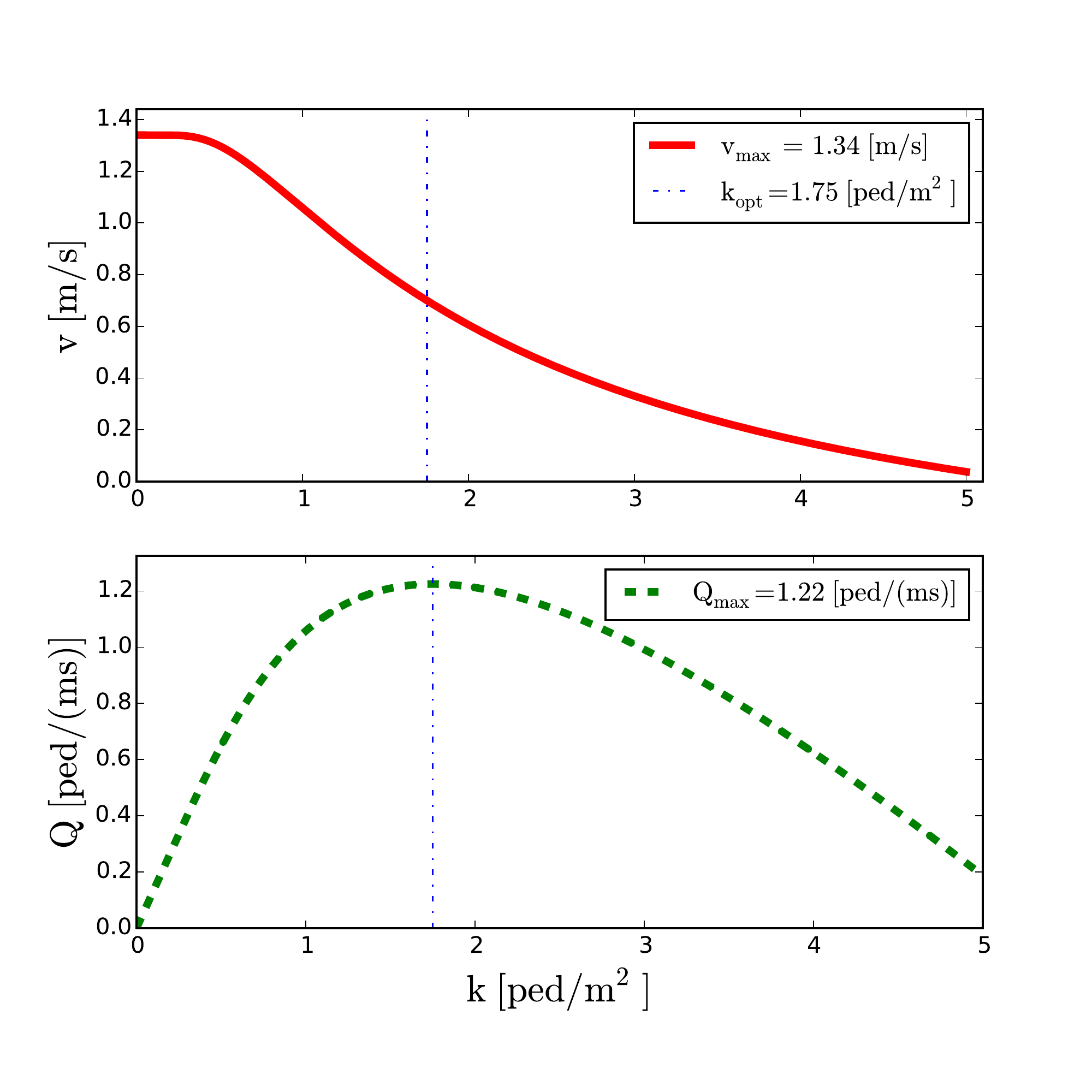}
    \put(0,88){(a)}
    \put(0,47){(b)}
    \end{overpic}

    \caption{\label{fig:fd} (a) Pedestrian density-velocity diagram \cite{weidmann1993transporttechnik} where the relationship between density $k$ and velocity $v = v_f(1.0-e^{-1.913(\frac{1.0}{k}-\frac{1.0}{k_{max}})})$. In this equation, the free-flow velocity $v_f = 1.34$ m/s. At maximum density $k_{max} = 5.4 \ \mathrm{ped/m^2}$, $v = 0$ m/s. (b) Pedestrian density-flow diagram \cite{weidmann1993transporttechnik} where the relationship between density $k$ and flow $Q(k) = k v(k) = kv_f(1.0-e^{-1.913(\frac{1.0}{k}-\frac{1.0}{k_{max}})})$. In this equation, the optimum density $k_{opt} = 1.75 \ \mathrm{ped/m^2}$ and the corresponding maximum flow $Q_{max} = 1.22$ ped/ms.}
    
\end{figure}

We incorporate Weidmann's fundamental diagram to describe pedestrian behaviour \cite{weidmann1993transporttechnik} shown in FIG.~\ref{fig:fd} into the model. Eq.~(\ref{eq:kv}) describes the relationship between density $k$ and velocity $v$. When $k = 0$ $\mathrm{ped/m^2}$, the free-flow velocity $v_f$ is $1.34$ m/s. At maximum density when $k_{max} = 5.4$ $\mathrm{ped/m^2}$, $v$ is $0$ m/s.

\begin{equation}
    v = v_f(1.0-e^{-1.913(\frac{1.0}{k}-\frac{1.0}{k_{max}})})
    \label{eq:kv}
\end{equation}

 The relationship between density $k$ and flow rate $Q$ follows as Eq.~(\ref{eq:kQ}). We can differentiate this equation to derive the optimum density $k_{opt} = 1.75$ $\mathrm{ped/m^2}$ when $dQ/dk = 0$. The corresponding maximum flow rate $Q_{max} = 1.22$ ped/ms.

\begin{equation}
    Q = k v(k) = k v_f(1.0-e^{-1.913(\frac{1.0}{k}-\frac{1.0}{k_{max}})})
    \label{eq:kQ}
\end{equation}

 As pointed out in the introduction, the fundamental diagram is not able to describe system dynamics far from equilibrium (i.e. high density crowds). Therefore, to ensure that agent movement can occur at high densities, we implement a network link density cap at $k_{cap} = 5 \ \mathrm{ped/m^2}$ such that minimum velocity is never less than $v_{min} = 0.04$ m/s. This ensures a minimum flow of $Q_{min} = 0.05 \ \mathrm{ped/(ms)}$ \cite{johansson2009constant}. Without this cap, as $k$ approaches $k_{max}$, $v$ tends towards $0$ m/s, which means that the simulations would run indefinitely.

\begin{figure}[!h]
  \centering
    \begin{overpic}[scale=0.4]{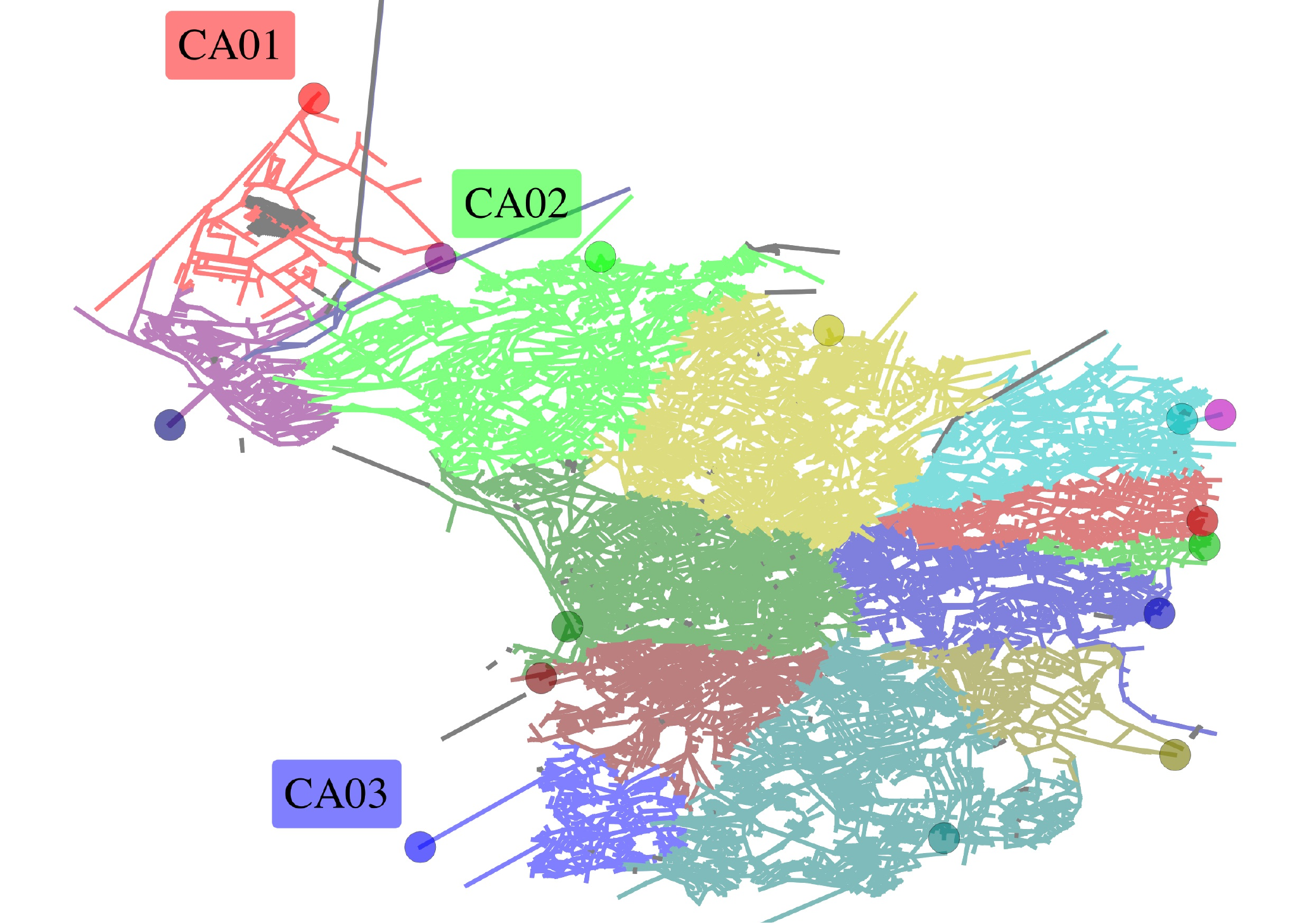}
    \put(0,65){(a)}
    \end{overpic}

    \begin{overpic}[scale=0.5]{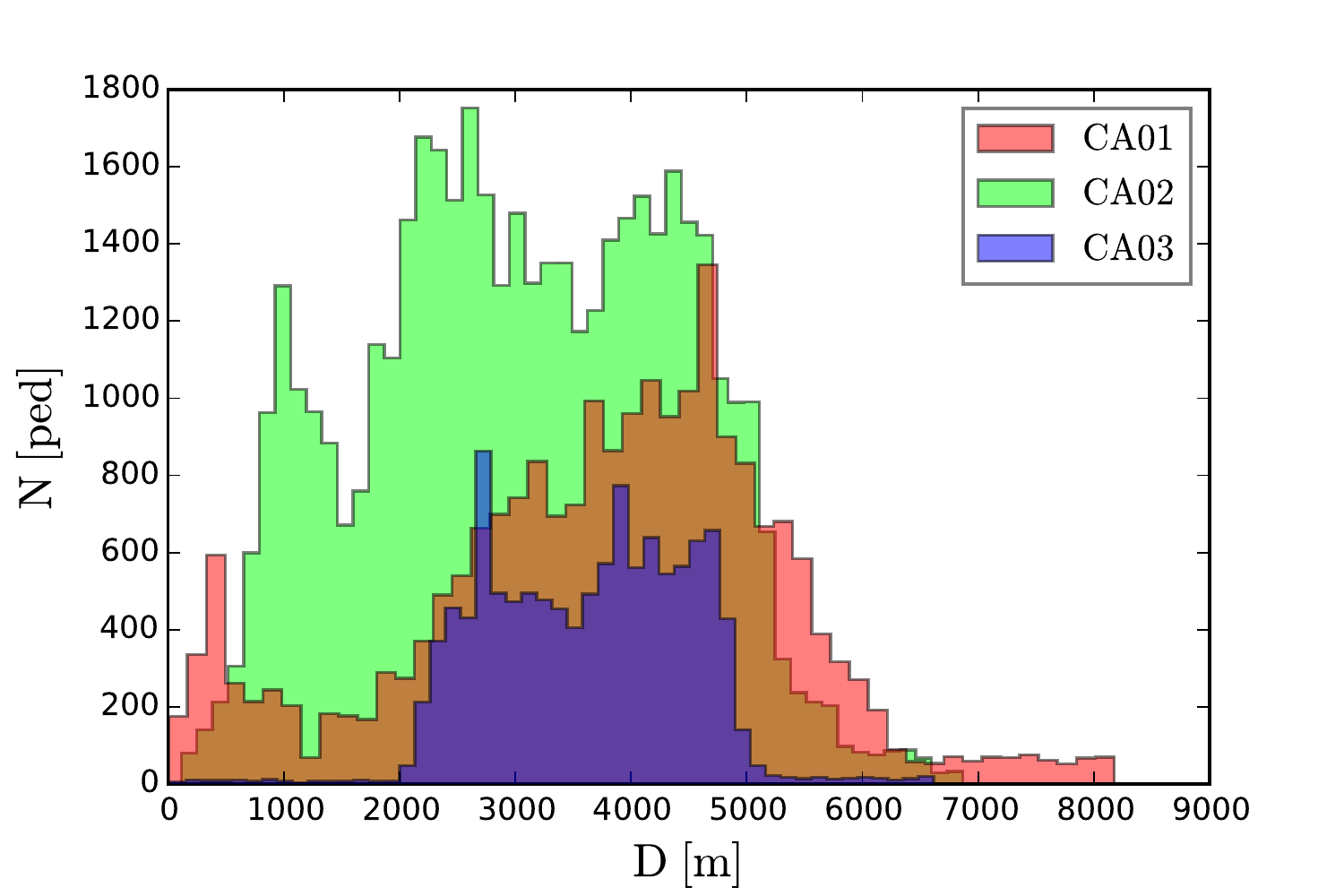}
    \put(-7,60){(b)}
    \end{overpic}

  \caption{\label{fig:catchment}(Colour Online) (a) Catchment areas (CA) are obtained by allocating agents to the exit node nearest to their initial position. For this, we use Dijkstra's shortest path algorithm \cite{dijkstra1959note}. Each colour in the figure represents one of the 15 City of Bristol CA. (b) Example showing distribution of agent distance to exit node $D$ for $3$ City of Bristol CAs denoted by red (CA01), green (CA02) and blue (CA03) histograms.}
\end{figure}

Under these assumptions, we select $50$ cities similar in area to City of Bristol ($235.82 \pm 25\%$ $\mathrm{km^2}$). We use network topology approximated from OpenStreetMap (OSM) \cite{haklay2008openstreetmap}. OSM is a source of Volunteered Geographic Information (VGI) \cite{goodchild2007citizens} growing in both contributor base and data quality \cite{haklay2010good,neis2012analysing,ali2014ambiguity}. We further divide these cities into $697$ catchment areas (CA), which we define as network components that emerge as agents are assigned to an exit node nearest to their initial position calculated using Dijkstra's shortest path algorithm \cite{dijkstra1959note}. FIG.~\ref{fig:catchment}a illustrates CA formation for City of Bristol and FIG.~\ref{fig:catchment}b shows how the distribution of initial agent distance to their exit node $D$ varies between different CAs.

\begin{figure}
    [!b] \centering

    \begin{overpic}[scale=0.6]{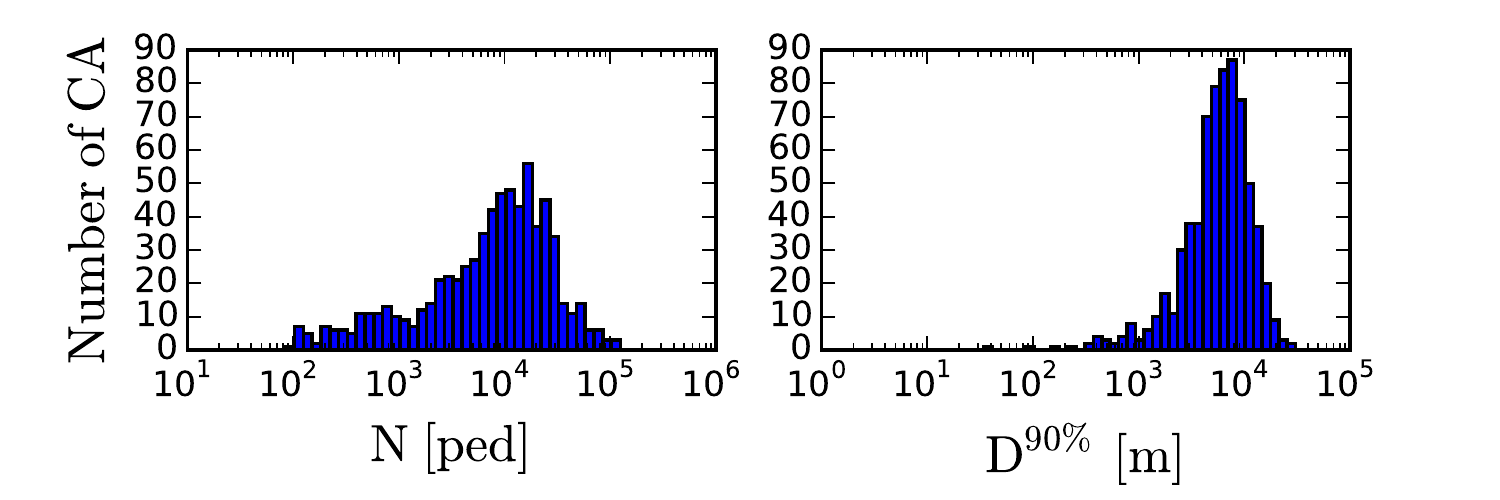}
    \put(8,35){(a)}
    \put(50,35){(b)}
    \end{overpic}

    \caption{(a) Histogram of catchment area (CA) populations $N$. (b) Histogram of agent distances to exits for $90\%$ of all CA agents $D^{90\%}$.}
    \label{fig:N-D90}    
\end{figure}

\newpage

\subsection{\label{sub:characteristic} Characteristic Variables}

\begin{figure}
    [!b] \centering

    \begin{overpic}[scale=0.55]{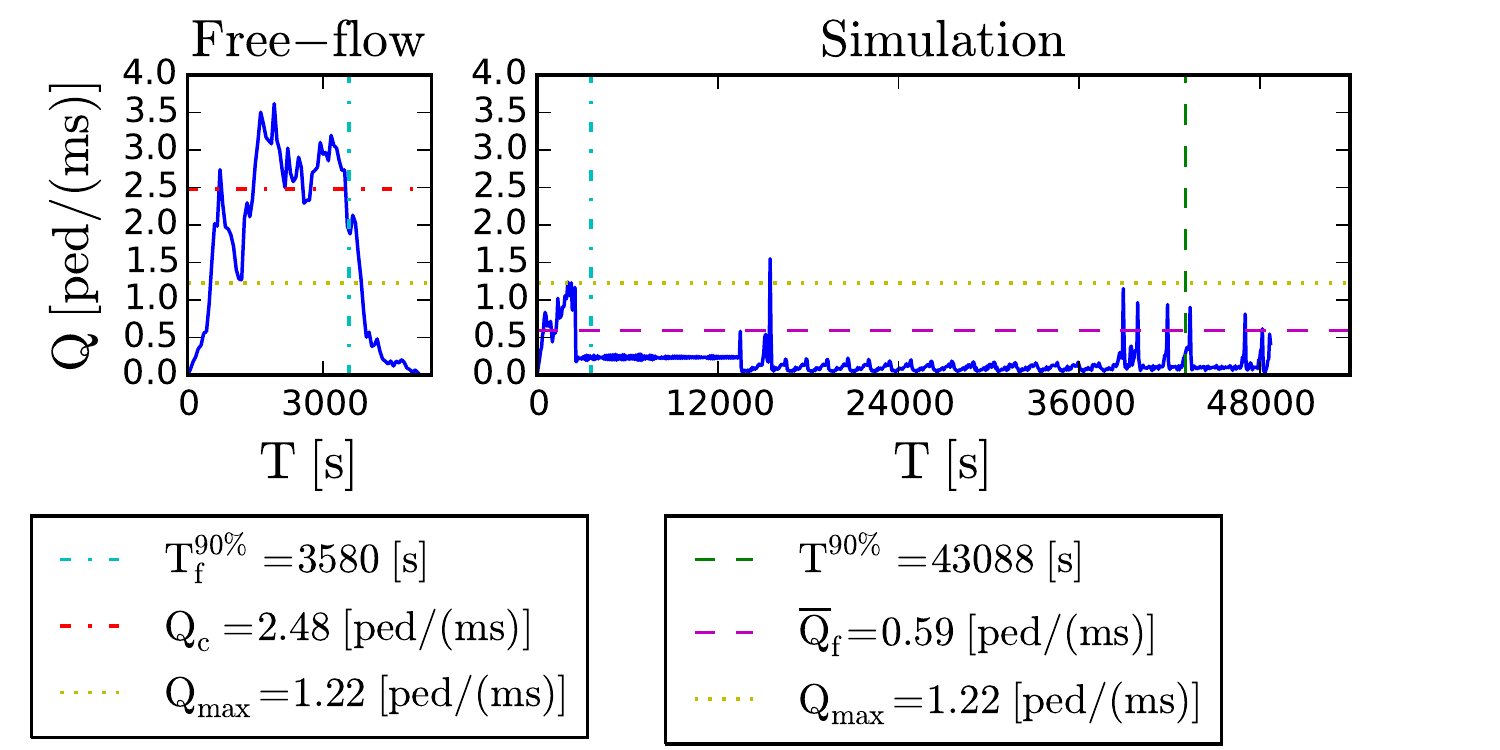}
    \put(5,50){(a)}
    \put(30,50){(b)}
    \end{overpic}

    \caption{\label{fig:single-T-Q} An example to show the position of free-flow time for $90\%$ of all catchment area (CA) agents $T^{90\%}_f$ and simulated time for $90\%$ of all CA agents $T^{90\%}$ using a City of Bristol CA (CA02). It also marks the position of maximum flow rate $Q_{max}$, characteristic free-flow time averaged flow rate $Q_c$, mean of simulated exit node flow rate in free-flow regime $\overline{Q}_f$.}
    
\end{figure}

Characteristic variables independent of dynamic agent interaction informs part of our analysis.
The first of these is the characteristic flow rate $Q_c$ described by Eq.~(\ref{eq:Qc}). It is defined as free-flow time averaged flow whereby we assume infinite link capacity. To illustrate the point, we mark the position of $Q_c$ for an example CA in FIG.~\ref{fig:single-T-Q}a.

\begin{equation}
    Q_c = \frac{N}{T^{90\%}_f W}
    \label{eq:Qc}
\end{equation}

We calculate $Q_c$ using CA population $N$, exit node width $W$ and free-flow catchment area traversal time for $90\%$ of all CA agents $T^{90\%}_f$, which is also our second characteristic variable.
We estimate $N$ from GRUMPv1 year $2000$ population dataset \citep{balk2005mapping} uniformly scaled up by a factor of $9.37\%$ in order to account for the rise in UK population between the years $2000$ and $2015$ \citep{office2011national}. It has a granularity of $1\ \mathrm{km^2}$. FIG.~\ref{fig:N-D90}a shows how $N$ is distributed in log scale. Values range from $10^2$ to $10^5$, peaking at $10^{3.82} \approx 6,628$ agents.
For exit nodes tagged `motorway' on OSM, we assume $W = 7.5 \ \mathrm{m}$ and for those tagged `trunk' or `primary', $W = 5 \ \mathrm{m}$  \cite{kunwar2014large}.
$T^{90\%}_f = D^{90\%}/v_f$, where $D^{90\%}$ is the distance to exit node for $90\%$ of all agents and free-flow velocity $v_f = 1.34$ m/s. $T^{90\%}_f$ is also marked in FIG.~\ref{fig:single-T-Q}a. If we ignore all congestion and bottleneck effects, it provides a lower bound estimate of evacuation time for $90\%$ of all CA agents.
We use $D^{90\%}$ because it approximates the size of a CA as a scalar without the weight of the last decile skewing the result. FIG.~\ref{fig:N-D90}b shows how $D^{90\%}$ is distributed across all CA in log scale with values ranging between $10^1$ to $10^5$ metres. It peaks at $10^{3.74} \approx 5,435$ metres, a distance belt within which 90\% of all CA agents are situated.

\begin{figure}
    [!b] \centering

    \includegraphics[width=0.5\textwidth]{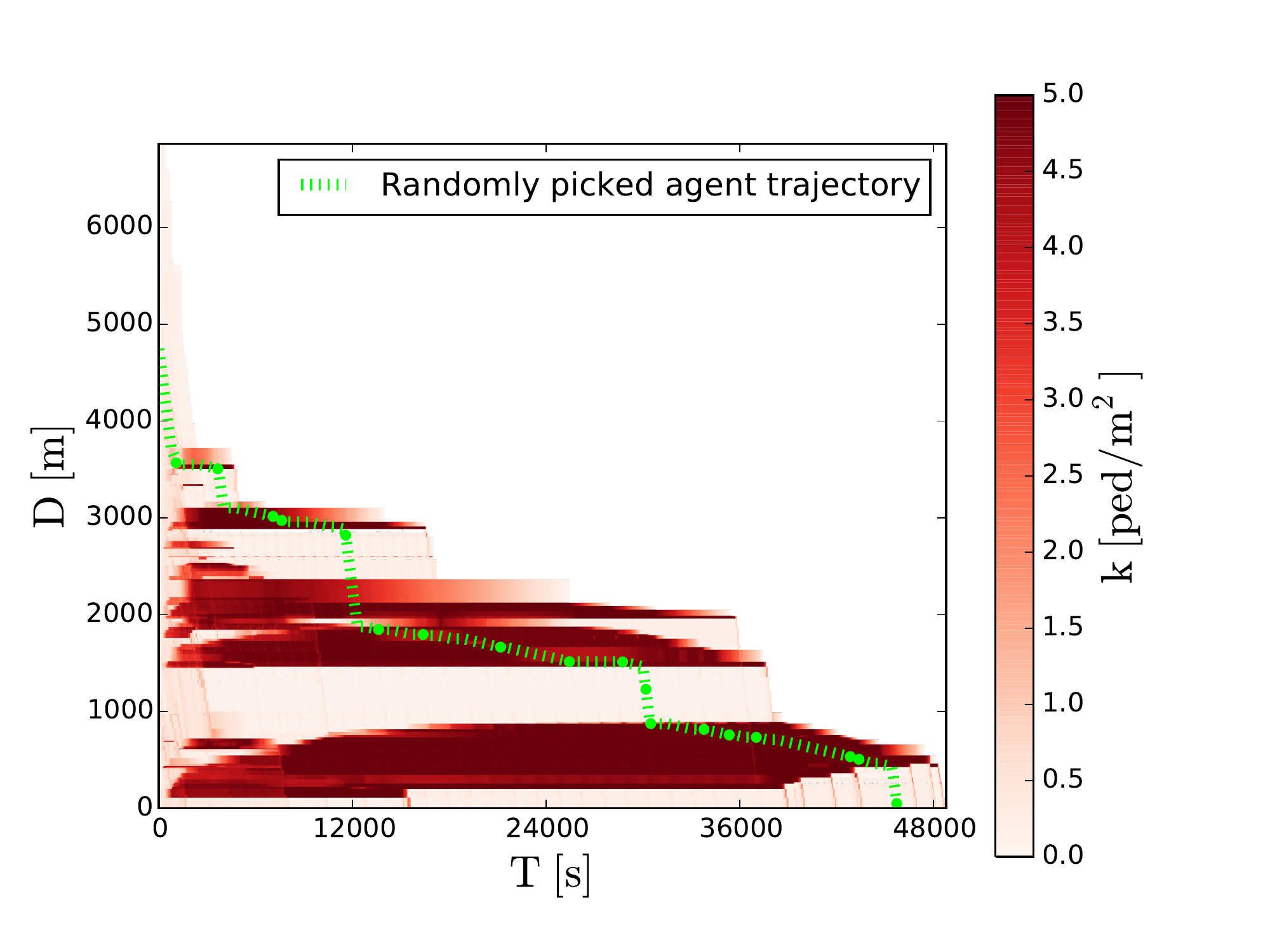}

    \caption{\label{fig:trajectory}Density $k$ at distance $D$ away from exit node at time $T$ where the density ranges between $0 \leq k \leq 5 \ \mathrm{ped/m^2}$ for a City of Bristol CA (CA02). Trajectory of a randomly picked agent is shown to illustrate how the agent velocity is reduced where the link density is high.}
    
\end{figure}

\subsection{\label{sub:simulated} Simulated Variables}

Simulated variables are obtained by studying dynamic interaction of agents under the following assumptions:

\begin{itemize}
    \item Agents immediately `walk' to the nearest exit on a signal to evacuate (i.e. pre-movement time is zero),
    \item Agents act independently (complex social behaviours such as family regrouping, co-operation, etc. are not taken into account).
\end{itemize}    

Our first simulated variable $\overline{Q}_f$ is defined as simulated exit node flow rate $Q$ averaged within the free-flow regime ($T<T^{90\%}_f$).
The area under the flow curve for each CA is proportional to the total number of agents. Larger this area before flow transitions to congested phase, bigger the $\overline{Q}_f$ value, precipitating a shorter congestion. Hence, the overall ETE is proportional to $\overline{Q}_f$.
We show the position of $\overline{Q}_f$ for an example CA in FIG.~\ref{fig:single-T-Q}b. The flow curve $Q$ it is derived from is calculated using Eq.~(\ref{eq:kQ}) where the density parameter $k = N/(WL)$, $N$ is the number of agents arriving at the exit node per time-step and $WL$ is the area of the exit link.
FIG.~\ref{fig:single-T-Q}b also marks the position of $T^{90\%}$, defined as the time at which $90\%$ of all CA agents arrive at the exit node. Unlike $T^{90\%}_f$, $T^{90\%}$ takes agent interaction and emergent bottlenecks into account.
While bottlenecks may be interspersed throughout a CA as shown by the example in FIG.~\ref{fig:trajectory} where  observed local density $k$ varies through distance from exit node $D$ and elapsed time $T$, it is ultimately the exit node flow rate $Q$ that influences the overall ETE. FIG.~\ref{fig:trajectory} also illustrates how velocity drops where density is high for a randomly picked agent trajectory.

\section{Linking Characteristic and Simulated Variables}
\label{results}

In this section, we establish the link between characteristic variable ($Q_c$, $T^{90\%}_f$) and simulated variables ($\overline{Q}_f$, $T^{90\%}$). We aggregate the simulated exit node flow $Q$ observed through absolute simulation time $T$. Then, we level the basis for comparison between CAs by normalising flow as $Q/Q_c$ and time as $T/T^{90\%}_f$. We substitute $Q$ and $T$ for $\overline{Q}_f$ and $T^{90\%}$ and define $\overline{Q}_f$ in relation to $Q_c$. We also define ratio $\overline{Q}_f/Q_c$ in relation to $T^{90\%}/T^{90\%}_f$. Using these relationships, we derive a general description of $T^{90\%}$ using characteristic variables $Q_c$ and $T^{90\%}_f$.

\begin{figure}
    [!htb] \centering

    \begin{overpic}[scale=0.58]{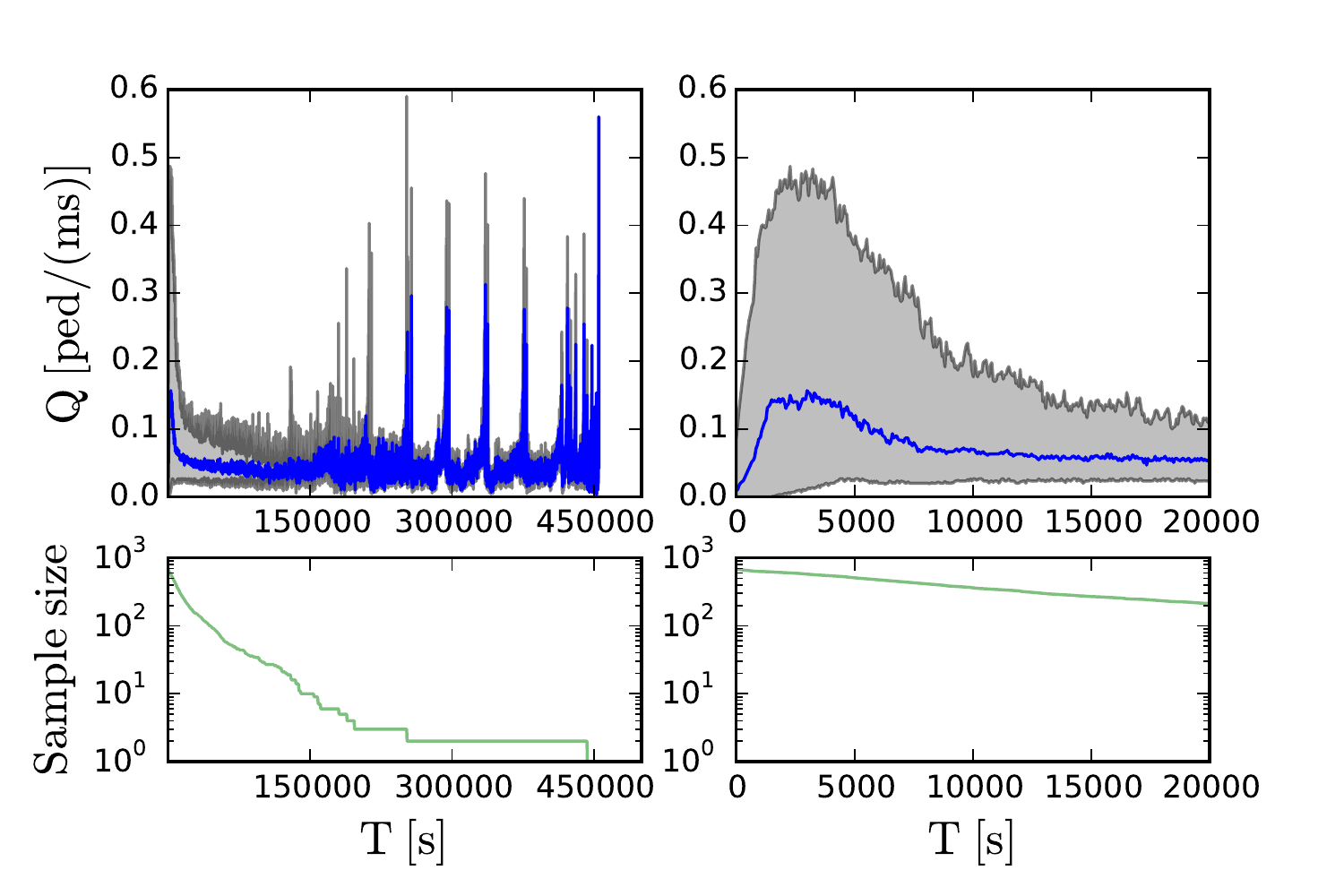}
    \put(0,62){(a)}
    \end{overpic}

    \begin{overpic}[scale=0.58]{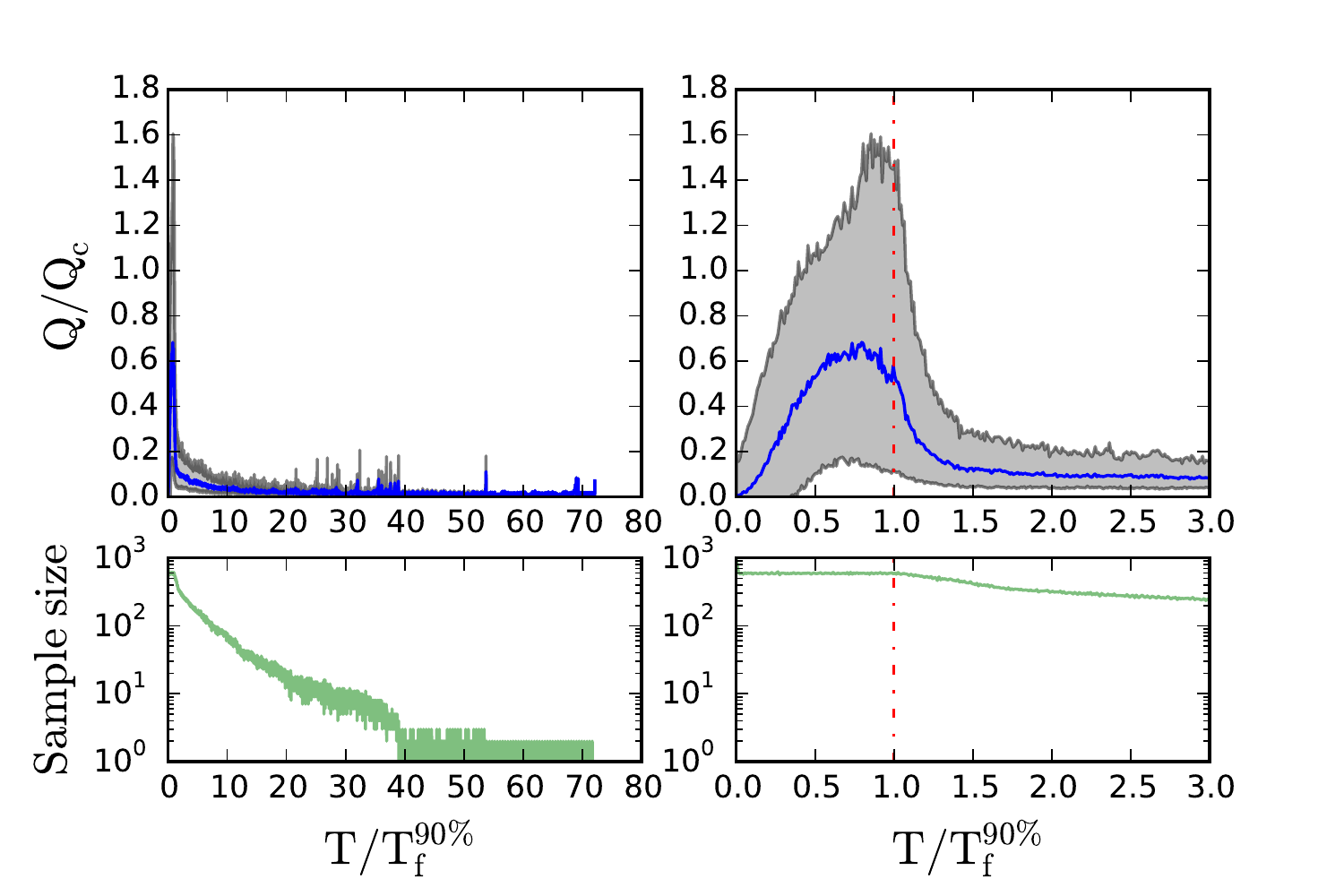}
    \put(0,62){(b)}
    \end{overpic}

    \caption{(a) Aggregate simulated flow at exit node $Q$ averaged across all CA over absolute simulation time $T$ where the grey region signifies $68\%$ confidence interval which becomes narrower with decreasing amount of aggregate data sample. (b) Aggregate exit node flow rate normalised by characteristic flow rate $Q/Q_c$ over time normalised by free-flow time for $90\%$ of all CA agents $T/T^{90\%}_f$ aggregated from all CA showing the $68\%$ confidence interval. $Q/Q_c$ peaks within $T/T^{90\%}_f < 1$ and mean $Q/Q_c < 1$.}
    \label{fig:all-T-Q}
\end{figure}

Aggregating simulated flow at exit node $Q$ across all CA over absolute simulation time $T$ produces FIG.~\ref{fig:all-T-Q}a. Looking at $0 < T < 20000$ band, we observe that the aggregate flows peak around $Q \approx 0.15$ ped/(ms) within a wide $68\%$ confidence interval early on in the simulation which gradually tapers. While the peak signals the transition from free-flow ($T \leq T^{90\%}_f$) to congested ($T > T^{90\%}_f$) regime, the exact point of transition is not clear in this representation. We also observe that as the sample size decreases with elapsing $T$, there is an increase in fluctuation of aggregate $Q$.
We normalise $T$ by $T^{90\%}_f$ and $Q$ by $Q_c$ to obtain FIG.~\ref{fig:all-T-Q}b. $Q/Q_c$ clearly peaks within $T/T^{90\%}_f < 1$ at $Q/Q_c \approx 0.6$ which implies that in general, $Q_c$ over-predicts the simulated flow. The flattening of the curve beyond the peak at $T/T^{90\%}_f \geq 1$ indicates the congested flow regime which carries on up to a maximum of $T/T^{90\%}_f \approx 72$. This is a significant gap between free-flow and simulated time but only applies to a small number of CA.

For the following part, we randomly divide our 697 CA into two datasets, the first half (the `training' dataset) to train our model with containing 347 CA and the second half (the `testing' dataset) to test our model containing 348 CA.

\begin{figure}
    [!t] \centering

    \begin{overpic}[scale=0.58]{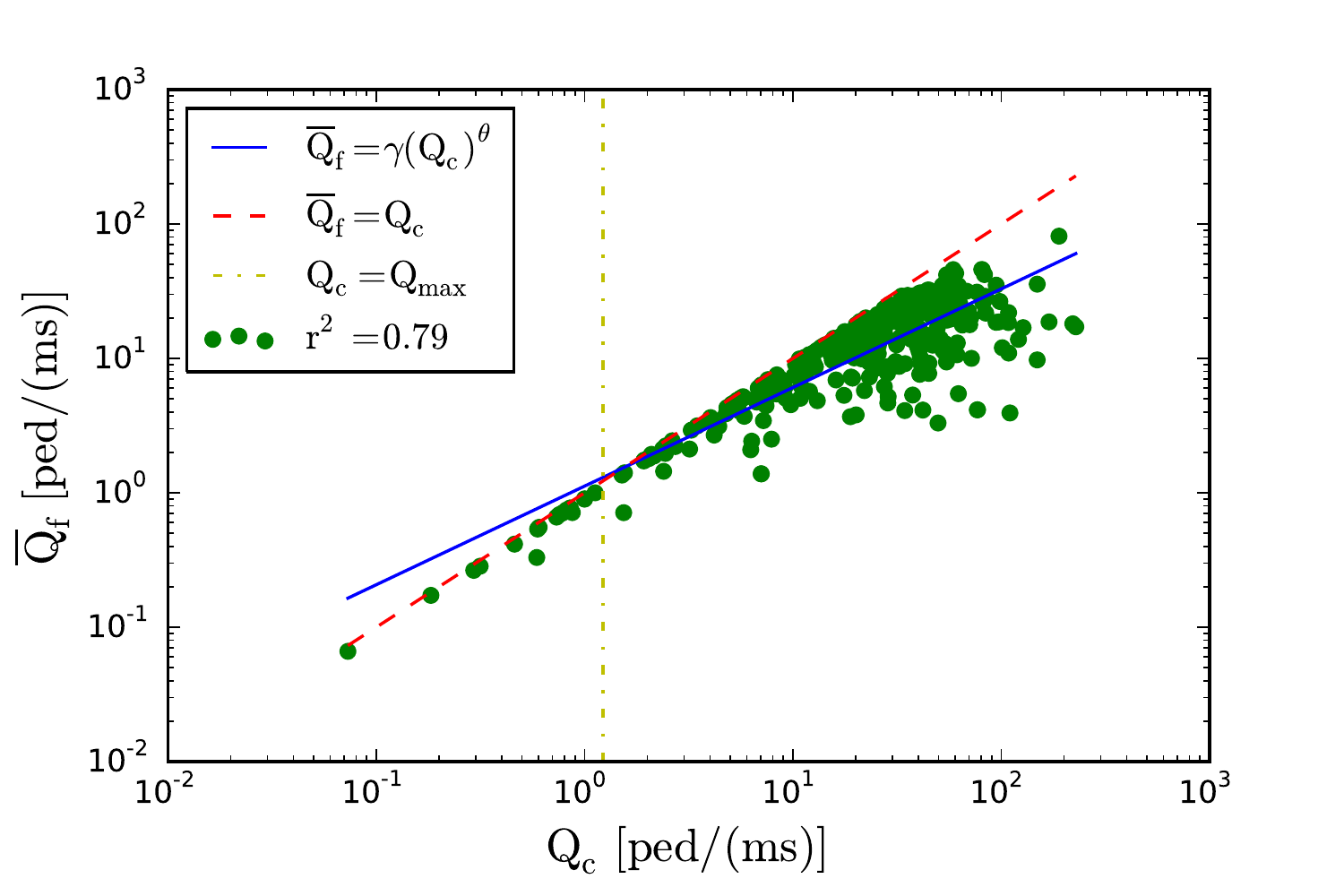}
    \put(0,62){(a)}
    \end{overpic}

    \begin{overpic}[scale=0.58]{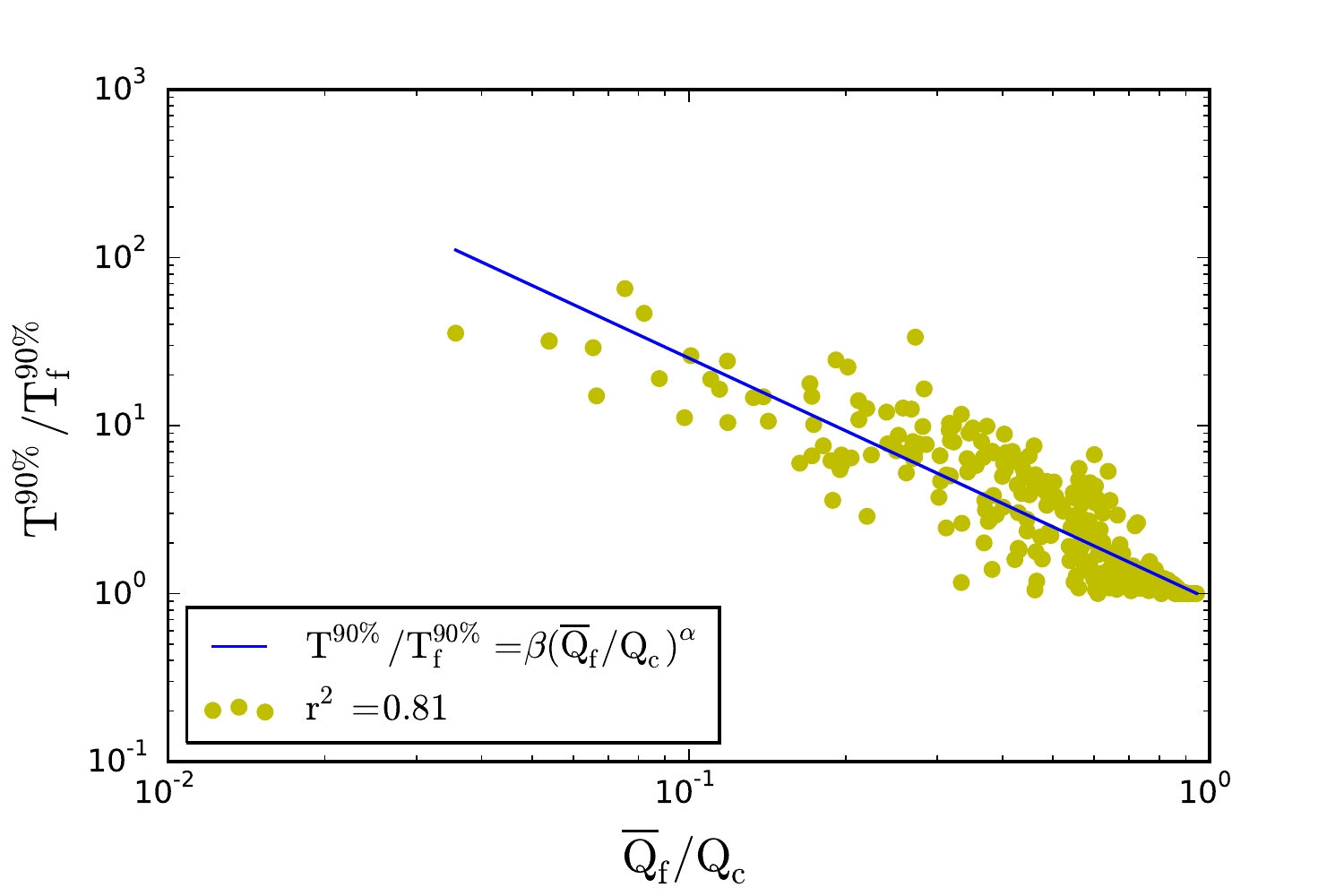}
    \put(0,62){(b)}
    \end{overpic}

    \caption{(a) Relationship between characteristic flow $Q_c$ and mean of simulated exit node flow rate in free-flow regime $\overline{Q}_f$. Each CA is represented by a data point. $\overline{Q}_f = Q_c$ is the upper bound. $\overline{Q}_f = \gamma (Q_c)^\theta$ describes the power-law best fit. (b) Relationship between the ratio of mean simulated exit node flow rate in free-flow regime to characteristic flow rate $\overline{Q}_f/Q_c$ to the ratio of simulated time to free-flow time for $90\%$ of all CA agents $T^{90\%}/T^{90\%}_f$. The best fit power-law equation is $T^{90\%}/T^{90\%}_f = \beta (\overline{Q}_f/Q_c)^\alpha$.}
    \label{fig:QT}
\end{figure}

Using the `training' dataset, we attempt to understand how $Q_c$ over-predicts simulated flow $\overline{Q}_f$. FIG.~\ref{fig:QT}a shows this relationship. The upper bound appears to be defined by $\overline{Q}_f = Q_c$ showing that $\overline{Q}_f$ never exceeds $Q_c$. There is a strong agreement between $Q_c$ and $\overline{Q}_f$ along the diagonal where $Q_c < Q_{max}$. However, when $Q_c > Q_{max}$, $\overline{Q}_f$ diverges from $\overline{Q}_f = Q_c$ line. It is better defined by a power-law fit ($r^2 = 0.79$) described by Eq.~(\ref{eq:Qmf}) where $\theta = 0.73$ and $\gamma = 1.12$.

\begin{equation}
	\overline{Q}_f = \gamma (Q_c)^\theta
	\label{eq:Qmf}
\end{equation}

We look for an equation to estimate the ETE, i.e. $T^{90\%}$ by analysing the relationship between ratios $\overline{Q}_f/Q_c$ and $T^{90\%}/T^{90\%}_f$ representing $Q$ and $T$ as $\overline{Q}_f$ and $T^{90\%}$ respectively.
$\overline{Q}_f/Q_c$ estimates the peak of the mean curve in FIG.~\ref{fig:all-T-Q}b.
For $T^{90\%}/T^{90\%}_f \gg 1$, delays due to agent interaction is  proportionately greater and as such, $T^{90\%}/T^{90\%}_f = 1$ is the best possible desired outcome.

We use the `training' dataset to derive the relationship seen in FIG.~\ref{fig:QT}b between $\overline{Q}_f/Q_c$ and $T^{90\%}/T^{90\%}_f$ with axes. There is a strong correlation ($r^2 = 0.81$) between the data points. For values of $\overline{Q}_f/Q_c \approx 1$, $T^{90\%}/T^{90\%}_f \approx 1$ implying that $\overline{Q}_f \approx Q_c$ when $T^{90\%} \approx T^{90\%}_f$. However, $\overline{Q}_f/Q_c \rightarrow 0$ as $T^{90\%}/T^{90\%}_f \rightarrow \infty$ since agents overflow into the congested regime. When $\overline{Q}_f \ll Q_c$, $T^{90\%} \gg T^{90\%}_f$.
The relationship between the two ratios is well described by the power-law of Eq.~(\ref{eq:T90-T90f}) with best fit parameter values $\alpha = -1.44$ and $\beta = 0.92$.

\begin{equation}
	\frac{T^{90\%}}{T^{90\%}_f} = \beta \Bigg(\frac{\overline{Q}_f}{Q_c}\Bigg)^\alpha
	\label{eq:T90-T90f}
\end{equation}

In order to obtain at least the first order estimate of ETE for a new CA without running an ABM simulation, we can equate $T^{90\%}$ solely in terms of characteristic variables $T^{90\%}_f$ and $Q_c$. We do this by substituting Eq.~(\ref{eq:Qmf}) into Eq.~(\ref{eq:T90-T90f}) to obtain Eq.~(\ref{eq:T90}) where $\phi = \alpha(\theta-1) = 0.38$ and $\omega = \beta \gamma ^{\alpha} = 0.78$.

\begin{equation}
	T^{90\%} = \omega (Q_c)^{\phi} T^{90\%}_f 
	\label{eq:T90}
\end{equation}

\begin{figure}
    [!htb] \centering

    \includegraphics[width=0.5
    \textwidth]{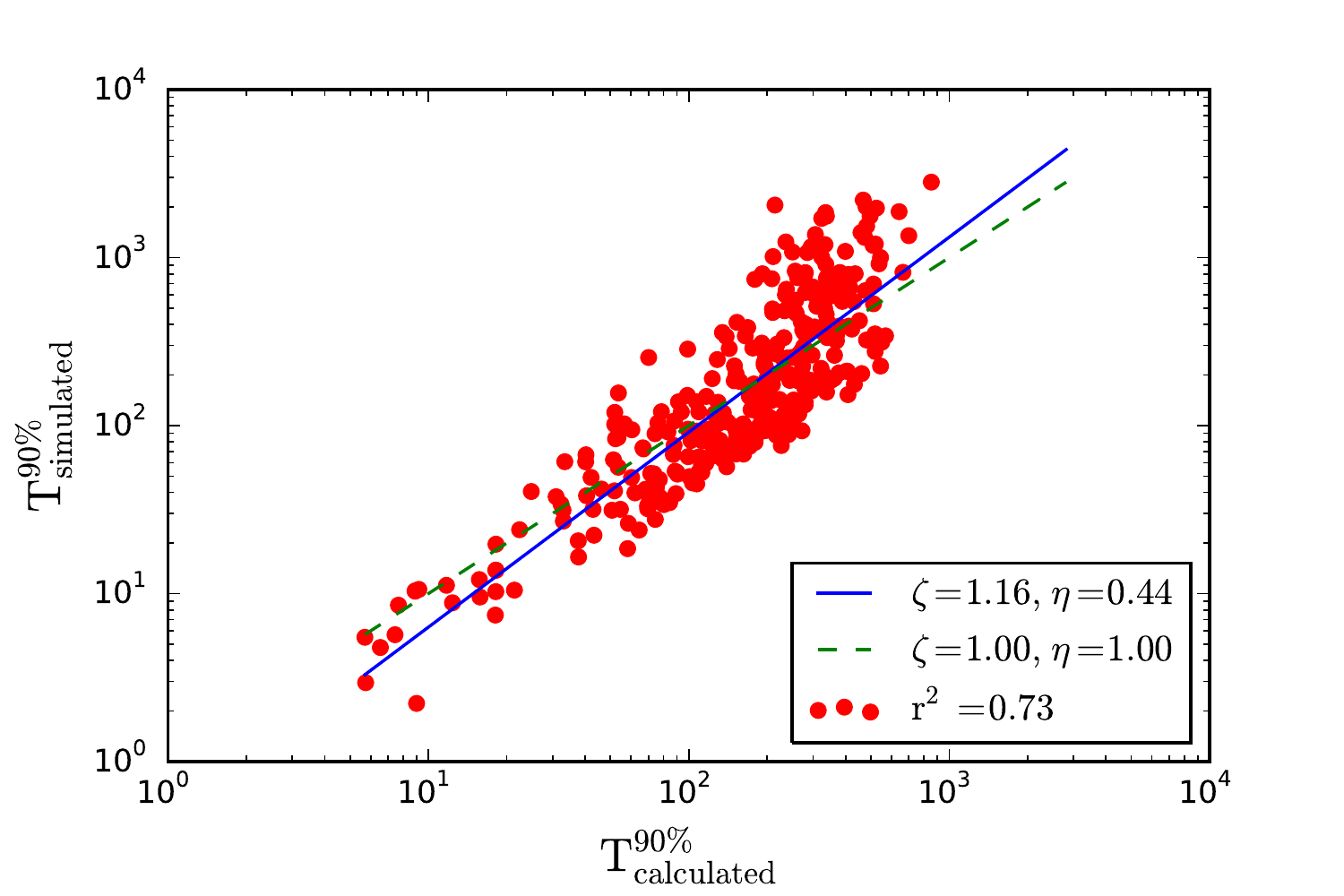}

    \caption{Comparing $T^{90\%}_{calculated}$ against $T^{90\%}_{simulated}$ where $T^{90\%}_{simulated} = \eta (T^{90\%}_{calculated})^\zeta$. When compared with mirror diagonal line ($\zeta = 1.00$, $\eta = 1.00$), the calculated values slightly underestimate the simulated values for higher values of $T^{90\%}$ ($\zeta = 1.16$, $\eta = 0.44$).}
    \label{fig:T90T90}
\end{figure}

We verify Eq.~(\ref{eq:T90}) using the `testing' dataset. We use $Q_c$ and $T^{90\%}_f$ parameters alone to calculate $T^{90\%}$ for each CA in this dataset and compare them against their simulated counterpart. According to FIG.~\ref{fig:T90T90}, there is a good agreement between the values ($r^2 = 0.73$) where $T^{90\%}_{simulated} = \eta (T^{90\%}_{calculated})^\zeta$. The calculated values under-estimate the simulated values (exponent $\zeta = 1.16$, coefficient $\eta = 0.44$) for higher values of $T^{90\%}$, as shown by the deviation of the trend from mirror diagonal line (exponent $\zeta = 1.00$, coefficient $\eta = 1.00$).


\section{\label{conclusion}Conclusions}

In conclusion, by exploring the underlying relationship between simulated ETE and $697$ CA attributes from $50$ UK cities, we present a method for calculating ETE, all using CA attributes: population, size and exit node width alone. This method more reliably estimates the ETE when characteristic flow rate is similar to mean of simulated exit node flow rate in the free-flow regime. There are discrepancies which exist between calculated and simulated ETE because statistical analyses do not fully capture the unique attributes of each CA. Hence, ABMs are better placed to deal with problems with many interacting entities. In our future work, we want to search for topological attributes that uniquely describe each CA which explain these discrepancies.

%

\begin{acknowledgments}
BK is grateful for funding from EPSRC Doctoral Training Grant and University of Bristol Systems Centre Open Innovation Industry Scholarship.
\end{acknowledgments}

\bibliography{library}

\end{document}